\documentstyle[aps,prd,epsfig]{revtex}
\begin{document}

\renewcommand{\topfraction}{1.0}
\twocolumn[\hsize\textwidth\columnwidth\hsize\csname
@twocolumnfalse\endcsname
\title{Heavy nuclei at the end of the cosmic ray spectrum?}
\author{L. A. Anchordoqui$^{1}$\thanks{Electronic address: 
doqui@venus.fisica.unlp.edu.ar}, 
G. E. Romero$^{2}$\thanks{Member of
CONICET}, and J. A. Combi$^{2 \dagger}$} 
\address{$^{1}$ Departamento de F\'{\i}sica, Universidad Nacional de La Plata, C.C.
67, 1900 La Plata, Argentina}
\address{$^{2}$ Instituto Argentino de Radioastronom\'{\i}a, C.C. 5, 1894 Villa
Elisa, Argentina}

\maketitle

\begin{abstract}

We provide an account of the possible acceleration of iron nuclei up to
energies $\sim300$ EeV in the nearby, metally-rich starburst galaxy NGC 253.
It is suggested that particles can escape from the nuclear region with
energies of $\sim10^{15}$ eV and then could be reaccelerated at the
terminal shock of the galactic superwind generated by the starburst,
avoiding in this way the photodisintegration expected if the nuclei were
accelerated in the central region of high photon density. We have
also made estimates of the expected arrival spectrum, which displays a
strong dependency with the energy cutoff at the source.   

\noindent {\it PACS number(s):} 96.40 - 98.70.S - 95.85.R  - 13.85.T

\end{abstract}

\vskip2pc]

\section{Introduction}

The discovery of extensive air shower events with energies
above 100 EeV (see Ref. \cite{YD} for a recent survey) confirms 
that the cosmic ray (CR) spectrum does not end with the 
expected Greisen-Zatsepin-Kuz'min (GZK) cutoff \cite{gzk}. The 
origin of this GZK cutoff is energy degradation of the CR particles
(usually assumed to be nucleons and nuclei) by
resonant scattering processes with the diffuse background radiation
that permeates the universe. The observed tail of the spectrum
could be, consequently, originated in a bunch of nearby
sources \cite{nosotros}.

Preferred sites for proton acceleration are 
astrophysical scenarios where large-scale shocks occur, as for instance in the hot spots of powerful
radio-galaxies \cite{R-B}. It could appear that, because of the
high-energy cutoff of shock acceleration increases with the charge number
of the nucleus, heavy ions would be nice candidates for ultra
high-energy
CRs. However, AGNs and radio galaxies are widely thought as regions of
very low  
metallicity and, in addition, it is well
established that above 200 EeV nuclei should be photodissociated 
by the 2.73 K photon background in a few Mpc \cite{esteban}. Thereupon,
nuclei acceleration up to the highest 
energies within astrophysical
environments is seldom considered in the literature.   
                             
Nonetheless, there has been a recently renewed interest in the propagation
of heavy 
nuclei \cite{esteban,sudaf,Ste97,prd2,Ste98}. This renewal is mainly
sustained by two facts: i) a medium
mass nucleus is the particle that provides the best fits of the shower
development of the highest energy CR event \cite{H}, and  ii) the arrival
direction of such event roughly points towards the nearby 
metally-rich galaxy M82 \cite{elbert-sommers}. 

Despite the aforementioned studies on nuclei propagation, it is far from
clear whether iron or other heavy nuclei can be accelerated up to energies
$\sim 300$ EeV in starburst galaxies like M82. One could naively expect
that, since the size scale of the starburst region is of the order of the
gyroradius of 300 EeV--($Z=26$) nuclei, strong shocks could diffusively
accelerate these ions to ultra-high energies. But this hope fades away as
soon as one notices the large photon energy densities (mostly in the far
infrared) measured in the central regions of these kind of galaxies: iron
nuclei are photodisintegrated long before they can reach the required
Lorentz factors.

We shall argue in this paper that, despite the mentioned problem, iron
nuclei can be actually acclerated in nearby starburst galaxies up to energies $\sim300$ EeV if a two-step process is
involved. The crucial point is that for energies above $\sim10^{15}$ eV,
acceleration occurrs in the terminal shock of the starburst superwind,
well outside the problematic nuclear region. Since NGC
253 is a southern object which has been scarcely discussed in relation
to CRs (for a brief discussion of M82 as CR accelerator 
see \cite{elbert-sommers}), 
we shall focus on it for our quantitative estimates. However, we emphasize
that due to the similarity between both galaxies, our conclusions
will be correct to M82 within the order of magnitude.  
Let us start with a recapitulation of some observational features of NGC 253.

\section{The starburst galaxy NGC 253}

Starbursts are galaxies undergoing a massive and large-scale star
formation episode. Their characteristic signatures are strong infrared
emission (originated in the high levels of interstellar extinction), a
very strong HII-region-type emission-line spectrum (due to a large number
of O and B-type stars), and a considerable radio emission produced by
recent supernova remnants. Typically, the starburst region is confined to
the central few hundreds of parsecs of the galaxy, a region that can be
easily 10 or more times brighter than the center of normal spiral galaxies.

NGC 253 has been described as the archetypal starburst galaxy by
Rieke {\it et
al.} \cite{rieke}, and as a prototype of superwind galaxy by Heckman and
collaborators \cite{heckman}. This object, whose distance is estimated in
the range 2.5 - 3.4 Mpc \cite{davidge,wynn}, has been extensively studied
from radio to $\gamma$-ray wavelengths \cite{beck,paglione,ptak}. More
than 60 idividual compact radio sources have been detected within the
central 200 pc of the nuclear region of NGC 253 \cite{ulvestad}, most of
which are supernova remnants (SNRs) of only a few hundred years old.
According to estimates from observations at different frequencies the
supernova rate is as high as $0.2-0.3$ yr$^{-1}$ \cite{ulvestad,forbes}. 

The central $\sim80$ pc of the galaxy contain around 24.000 O and 3.000
red supergiant stars \cite{forbes}, in addition to the SNRs and numerous
HII regions. This means
that the massive star formation rate is $\sim 0.1$ M$_{\odot}$ yr$^{-1}$.
Strong [Fe II] emission has been also detected with a total [Fe II] (1.644
$\mu$m) luminosity of $\sim2.8\times10^{39}$ erg s$^{-1}$, which reflects
the very rich iron production in the supernovae and their associated
shocks \cite{engelbracht}.

In the light of such a concentrated activity it is not surprising that
strong physical, morphological, and kinematic evidence for the existence
of a galactic superwind has 
been found in NGC 253 \cite{mccarthy,heckman}. Galactic-scale superwinds are
driven by the collective effect of supernovae and massive star winds. The
high supernovae rate creates a cavity of hot gas ($\sim10^8$ K) whose
cooling time is much greater than the expansion time scale. Since the wind
is sufficiently powerful, it can blow out the interstellar medium of the
galaxy avoiding to remain trapped as a hot bubble. As the cavity
expands a strong shock front is formed on the contact surface with the
cool interestellar medium. Shock interactions with low and high density
clouds produce X-ray continuum and optical line emission, respectively,
that has been directly observed \cite{mccarthy}. In addition, kiloparsec
regions well outside the disk present double emission-line profiles with
line splitting of 200-600 km s$^{-1}$, a clear evidence of mass motion.
The morphology of the optical emission line nebulae indicates that the
outflowing gas is located along the walls of a cone that is
limb-brightened, typical of a superwind in a blowout phase.  
 
The shock velocity can reach several thousands
of kilometers per second and ions like iron nuclei can be then efficiently
accelerated in this scenario up to high energies ($\sim 10^{20}$ eV) by Fermi mechanism as
we shall discuss in the next section.

\section{CR acceleration at NGC 253}

We suggest that the iron nuclei acceleration in NGC 253 occurs through a
two-step process. In a first stage, ions are diffusively accelerated  at
single supernova shock waves within the nuclear region of the galaxy.
Energies up to $\sim 10^{15}$ eV can be achieved in this step \cite{lagage}.
Fe-nuclei are not photodissociated in the process despite the starburst's
central photon density is much larger than that of the Milky Way. The
continuum spectrum of NGC peaks in the far infrared at $\sim100\mu$m, with
a luminosity of $\sim 3\times 10^{10}$ $L_{\odot}$ \cite{rice} and a photon
energy density of $U_{\rm ph}\sim 200$ eV cm$^{-3}$ \cite{heckman}.
For such values is straightforward that the nuclei interactions with
the blue--shifted ambient photons are quite below the
photodisintegration threshold. Energy losses through pair production are
also negligible.
  
By other hand, interactions with the rich interstellar gas 
in the center of the galaxy might disintegrate the ions if the escape from
the starburst region is dominated by diffussion. Multiwavelength molecular
line observations of the central region show that the average density of
the molecular clouds is $\sim 10^5$ cm$^{-3}$, with a filling factor
$<10^{-3}$ \cite{new}. This means an average gas density (mainly H$_2$) in
the active region within the range 30 - 300 cm$^{-3}$ \cite{paglione}.
The cross section for Fe$-$H$_2$
interactions ($\sigma \approx 1463.7$ mb) can be estimated from Rudstam's
parametrization 
of proton induced spallation, $\sigma = 50 A_t^{2/3}$ mb 
(where $A_t$ is the mass number of the target
nucleus) \cite{rudstam}. The mean free path for the iron nuclei is then 
$\lambda \sim (n\sigma_{\rm Fe-H_2})^{-1}$ which results in the 
range   738 - 7380 pc when the upper and lower limits of particle
density are considered. The central starburst region can be modeled as a
disk of 70 pc thick with a radius $R \approx 300$ pc \cite{paglione}.
Since the gyroradius of the Fe nuclei is $\sim 10^{-3}$ pc, they certainly
cannot be driven out from the starburst by diffusion.

However, due to the nature of the central region in NGC 253, the escape of
the iron nuclei is expected to be dominated by convection. In fact, the
presence of several tens of young SNRs with very high expansion velocities
($\sim 12000$ km s$^{-1}$ \cite{new2}) and thousands of massive O stars
(with stellar winds of terminal velocities up to 3000 km s$^{-1}$) must
generate collective plasma motions of several thousands of km per second.
Then, due to the coupling of the magnetic field to the hot plasma, the
magnetic field is also lifted outwards and forces the cosmic ray gas to
stream along from the starburst region.

The relative importance of convection and diffusion in the escape of the
cosmic rays from a region of disk scale height $h$ is given by the
dimensionless parameter,
\begin{equation}
q=\frac{V_0\,h}{\kappa_0},
\end{equation}
where $V_0$ is the convection velocity and $\kappa_0$ is the cosmic ray
diffusion coefficient inside the starburst \cite{new3}. When $q \alt 1$,
the cosmic ray outflow is ``difussion dominated'', whereas when $q \agt 1$
it is ``convection dominated''. Assuming for the central region of NGC 253 a
convection velocity of the order of the expanding SNR shells (i.e. $\sim$
10000 km s$^{-1}$, a scale height $h \sim 35$ pc, and a reasonable value
for the diffusion coefficient $\kappa_0 \sim 5 \times 10^{26}$ cm$^2$ s$^{-1}$
\cite{berezinsky-book}, we get $q \sim 216$ and convection dominates the
escape of the particles. The residence time of the iron nuclei in the
starburst results $t_{\rm RES} \sim h / V_0 \approx 1 \times 10^{11}$ s.
Most of the nuclei then escape through the disk in opposite directions
along the symmetry axis of the system, being the total path traveled
substantially shorter than the mean free path.

Once the nuclei escape from the central region of the galaxy with energies
of $\sim 10^{15}$ eV, they are injected into the galactic-scale wind and
experience further acceleration at its terminal shock.\footnote{CR
acceleration at superwind shocks was firstly proposed by Jokipii and Morfill
during the 80s \cite{bebito} in the context of our own Galaxy.} The scale length of
this shock is of the order of several tens of kpc (see Ref.
\cite{heckman}), so it can be considered as locally plane for
calculations. The shock velocity $v_{\rm sh}$ can be estimated from the
empirically determined superwind kinetic energy flux $\dot{E}_{\rm sw}$
and the mass flux $\dot{M}$ generated by the starburst through:

\begin{equation}
\dot{E}_{\rm sw}=\frac{1}{2} \dot{M} v_{\rm sh}^2.
\end{equation}

The shock radius can be approximated by $r\approx v_{\rm sh} \tau$, where
$\tau$ is the starburst age. Since the age is about a few tens of million
years, the maximum energy attainable in this configuration is constrained by
the limited acceleration time arisen from the finite shock's lifetime. For
this second step in the acceleration process, the photon field energy
density drops to values of the order of the cosmic background radiation
(we are now far from the starburst region), and consequently, as we
shall discuss in the next section, iron nuclei
are again safe from photodissociation while energy increases from $\sim
10^{15}$ to $10^{20}$ eV.

In order to estimate the maximum energy that can be reached by the nuclei
in the second stage of the acceleration process let us consider the
superwind terminal shock propagating in a homogeneous medium with an
average magnetic field $B$. If we work in the frame where the shock is at
rest, the upstream flow velocity will be ${\bf v_1}$ ($|{\bf v_1}|=v_{\rm
sh}$) and the downstream velocity, ${\bf v_2}$. The magnetic field
turbulence is assumed to lead to isotropization and consequent diffusion
of energetic particles which then propagate according to the standard
transport theory \cite{jokipii}. The acceleration time scale is then
\cite{drury}: 

\begin{equation}
t_{\rm acc}=\frac{4 \kappa}{v_1^2}  \label{t}
\end{equation}
where $\kappa$ is the upstream diffusion coefficient which can be written
in terms of perpendicular and parallel components to the magnetic field,
and the angle $\theta$ between the (upstream) magnetic field and the
direction of the shock propagation:

\begin{equation}
\kappa=\kappa_{\parallel} \cos^2\theta + \kappa_{\perp} \sin^2\theta
\end{equation}
   
Since strong turbulence is expected from the shock we can take the Bohm
limit for the upstream diffusion coefficient parallel to the field, i.e.

\begin{equation}
\kappa_{\parallel}=\frac{1}{3}\frac{E}{ZeB_1}
\end{equation}
where $B_1$ is the strength of the pre-shock magnetic field and $E$ is the
energy of the $Z$-ion. For the $\kappa_{\perp}$ component we shall assume,
following Biermann \cite{birmanncr1}, that the free mean path
perpendicular to the magnetic field is independent of the energy and has
the scale of the thickness of the shocked layer ($r/3$). Then,

\begin{equation}
\kappa_{\perp}=\frac{1}{3} r (v_1-v_2)
\end{equation}
or, in the strong shock limit,

\begin{equation}
\kappa_{\perp}=\frac{r v_1^2}{12}.
\end{equation}

Since the upstream time scale is $t_{\rm acc}\sim r/(3 v_1)$, we rewrite
Eq. (\ref{t}) as:

\begin{equation}
\frac{r}{3v_1}=\frac{4}{v_1^2}\left(\frac{E}{3ZeB_1} \cos^2\theta + \frac{r v_1^2}{12}
\sin^2\theta\right), 
\end{equation}
and then, using $r=v_1\tau$ and transforming to the observer's frame, we get:

\begin{equation}
E_{\rm max}\approx\frac{1}{4} ZeB v_{\rm sh}^2 \tau,
\end{equation}
or
\begin{equation}
E_{\rm max}\approx\frac{1}{2} ZeB \frac{\dot{E}_{\rm sw}}{\dot{M}} \tau,
\end{equation}
in terms of parameters that can be determined from observations.

The
predicted kinetic energy and mass fluxes of the starburst of NGC 253
derived from the measured IR luminosity are $2\times10^{42}$ erg s$^{-1}$
and 1.2 M$_{\odot}$ yr$^{-1}$, respectively \cite{heckman}. The starburst
age is estimated from numerical models that use theoretical evolutionary
tracks for individual stars and make sums over the entire stellar
population at each time in order to produce the galaxy luminosity as a
function of time \cite{edades}. Fitting the observational data these
models provide a range of suitable ages for the starburst phase that, in
the case of NGC 253, goes from $5\times 10^7$ to $1.6\times 10^8$ yr (also
valid for M82) \cite{edades}. These models must assume a given initial
mass function (IMF), which usually is taken to be a power-law with a variety of
slopes. Recent studies has shown that the same IMF can account for the
properties of both NGC 253 and M82 \cite{engelbracht}. Here we shall
assume a conservative age $\tau=50$ Myr.
Finally, the radio and $\gamma$-ray emission from NGC
253 are well matched by models with $B\sim50\mu$ G \cite{paglione}. With
these figures we obtain a maximum energy for iron nuclei of:

\begin{equation}
E_{\rm max}^{\rm Fe}\sim 3.4\times 10^{20}\;\;\;\;{\rm eV},
\end{equation}
a value quite lower (more than two orders of magnitude) than the
limit imposed by the synchrotron losses \cite{ray}.

\begin{figure}
\label{ngcf1}
\centering
\leavevmode\epsfysize=5.5cm \epsfbox{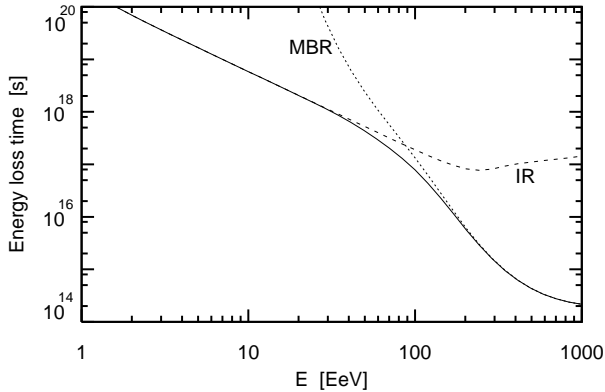}\\
\caption{Effective energy loss time for iron nuclei 
photodisintegration through the 2.73K and IR photons.
Adapted from Epele and Roulet Ref. [7].}
\end{figure}

Regarding the spectral slope of the particles, it is expected that they
emerge from the central region of NGC 253, where there are strong stellar
winds from the massive star population, with an energy index $\gamma=2.4$
\cite{birmanncr1}, which
is in fine accordance with the observed radio spectrum $\propto \nu^{-0.7}$
observed in the individual SNRs within the starburst \cite{ulvestad}. After
the
particles left the nuclear region, diffusive losses tend to steepen
the spectrum  at the terminal shock of the galactic superwind,
where reacceleration takes place. The final index will
depend on some not well known parameters as the Mach number of the
terminal shock. In what follows, we shall study the propagation of the
nuclei from NGC 253 to the Earth in models with $\gamma=2.4$, 2.5, and 2.6.

\section{Propagation effects}

The basic interactions between the universal background radiation and
nuclei of extremely high energy have been first discussed in detail
by Puget {\it et al} \cite{PSB}.
Heavy nuclei with energies above a few EeV get attenuated mainly
by photodisintegration off the microwave background radiation (MBR) and the
intergalactic infrared (IR) background photons.\footnote{Actually, the
pair creation process due to interactions with the MBR as well as 
disintegration with the optical background photons also attenuate the
nuclei energy. However, these processes are not essential for the
discussion  presented in this paper.
For details of these interactions the reader is referred to Ref. \cite{esteban}.}

The photodisintegration process is dominated by a broad maximum
designated as the giant dipole resonance which peaks in the
$\gamma$-ray energy range of 10 to 30 MeV (nucleus rest frame).
In the initial
absorption process, the photon energy may be given to a single
nucleon (although almost always is  shared with others), so the decay 
process might involve the emission of one (or more) nucleons.
The disintegration rate of a nucleus of mass $A$ with the subsequent 
production of $i$ nucleons reads \cite{Ste69},
\begin{equation}
R_{Ai} = \frac{1}{2 \Gamma^2} \int_0^{\infty} dw \,
\frac{n(w)}{w^2} \, \int_0^{2\Gamma w} dw_r
 \, w_r \sigma_{Ai}(w_r)
\label{rate}
\end{equation}
where $n(w)$ is the density of photons with energy $w$ in the 
system of reference in which the MBR is at 2.73 K.
In the formula, $w_r$ is the energy of the photons in the rest frame of the nucleus,
$\Gamma$ is the Lorentz factor, and $\sigma_{Ai}$
is the cross section for the interaction.
The total cross section of photonuclear interaction is well known
\cite{sol}. As usual, the MBR is described
by Plankian spectrum of 2.73 K. For the diffuse IR radio 
background (photon energies between $2 \times 10^{-3}$ - $8 \times 10^{-1}$ eV) 
we shall use the estimate given in Ref. \cite{IR},
\begin{equation}
\frac{dn}{dw} = 1.1 \times 10^{-4} \left(\frac{w}{{\rm
eV}}\right)^{-2.5} \,\, {\rm cm}^{-3}{\rm eV}^{-1}.
\end{equation}
With these figures at hand it is straightforward
to compute the energy loss time of iron nuclei (displayed in Fig. 1).
It is clear from these calculations that the universal background radiation
would not affect iron acceleration in NGC 253 up to a few thousands of EeV.
Moreover, since the nearness of the starburst galaxy, just the
interaction with the MBR becomes relevant in their travel to Earth.
The fractional energy loss as a function of the MBR energy
(Lorentz factor) have been already parametrized \cite{prd2},
\begin{mathletters}
\begin{equation}
R(\Gamma)=3.25 \times 10^{-6}\, 
\Gamma^{-0.643}                                          
\exp (-2.15 \times 10^{10}/\Gamma)\,\, {\rm s}^{-1} 
\label{par1}
\end{equation}
if $\Gamma \,\in \, [1. \times 10^{9}, 3.68 \times 10^{10}]$, and 
\begin{equation}
R(\Gamma) =1.59 \times 10^{-12} \, 
\Gamma^{-0.0698}\,\, {\rm s}^{-1}   
\label{par2}
\end{equation}
if $ \Gamma\, 
\in\, 
[3.68 \times 10^{10}, 1. \times 10^{11}]$. 
\end{mathletters}
It is noteworthy that the possible disintegration histories computed using
Eqs. (\ref{par1}) and (\ref{par2}) are in very good
agreement with Monte Carlo simulations when the sources are
located near the Earth (distances $\alt$ 10 Mpc) \cite{phd}.

Using the formalism sketched in \cite{prd2}, it is easily 
obtained the evolution of the differential spectrum
$Q(E_g,t) = K E_g^{-\gamma} \delta(t-t_0)$ of
the iron nuclei injected by NGC 253 at $t_0$.
The number of surviving fragments with energy $E$ at time $t$ 
is given by,
\begin{equation}
N(E, t) dE  = \frac{K E_g^{-\gamma+1}}{E} dE, 
\label{espectro}
\end{equation}
where $E_g$ denotes the energy at which the nuclei were 
emitted from the source, related to the energy detected on Earth by 
$E = E_g \,\, e^{-R(\Gamma ) \, t / 56}$ (recall that 
the Lorentz factor, $\Gamma = E_g / 56$, does not result modified
during the propagation). \footnote{Notice that the ($Z, A$)-dependence
of $R$ is roughly cancelled by dividing by $A$ in the exponent. This
implies  that $R$ can in fact be integrated down to lower, spallated $A$
values and still be reasonably accurate \cite{PSB}.}

\begin{figure}
\label{ngcf2}
\centering
\leavevmode\epsfysize=8cm \epsfbox{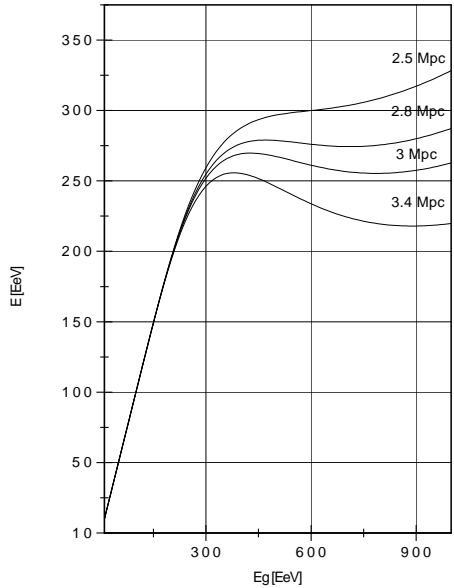}\\
\caption{Relation between the injection and arrival energies for
possible propagation distances from NGC 253 according to current data.}
\end{figure}

In Fig. 2 it is shown the energy degradation for different flying
timescales arising from different propagation regimes. The energy
spectrum of the surviving fragments is degenerate. For instance, for
a propagation distance of 3 Mpc, the composition of the arrival 
nuclei changes from $A=55$ (for $\Gamma \approx 3 \times 10^9$) to $A=44$ 
(for $\Gamma \approx 6 \times 10^{9}$). If a maximum energy of 340
EeV is attainable in the source, the relation between the injection
and arrival energies is a monotonously decreasing function.
At this stage, we conveniently introduce the modification factor $\eta$, 
defined as the ratio between the modified spectrum and the unmodified one. 
Notice that once the nuclei have energy enough to undergo
photodisintegration through the giant dipole resonance, the value of
the modification factor is always less than one.

Contrariwise, if the injection energy has an upper cutoff at $E \agt
560$ EeV (this can be achieved, for instance, with a higher value of $\tau$), 
the function which relates the injection and arrival
energies becomes multivalued, yielding a bump-like feature in the
modification factor. To understand this behavior, recall that nuclei
suffer a violent disintegration at these energies (see Fig. 1), in
such a way that, for a Lorentz factor $\Gamma \approx 8 \times 10^{9}$, the
composition of the arrival nuclei drops to $A \approx 33$. So,
particles injected with different energies might arrive with the same
energy, piling up around 250 EeV, just before the expected cutoff. The whole
effect can be clearly appreciated in Fig. 3, where we have plotted the
modification factor for two different cutoffs in the injection spectrum.

For longer propagation distances, the pile--up shifts to lower
energies (e.g. to $\approx 200$ EeV for 3.4 Mpc), broadening its profile.
On the other hand, for shorter distances to the nuclei--emitting source
there is no spectral bump, i.e., the spectrum is monotonic. Thus,
the probability of detecting events diminishes with increasing
Lorentz factor.

We remark that the change of the spectral index does not significantly affect
the shape of the modification factor. The height of the pile--up
diminishes with the steepening  of the injected spectrum, but this
effect is accompanied by a simultaneous drop in the CR flux of the
preceeding bin of energy, resulting in the same overall shape for
$\eta$, although downshifted (see Figs. 4 and 5 of ref \cite{prd2}).

\begin{figure}
\label{bwngc2}
\centering
\leavevmode\epsfysize=6cm \epsfbox{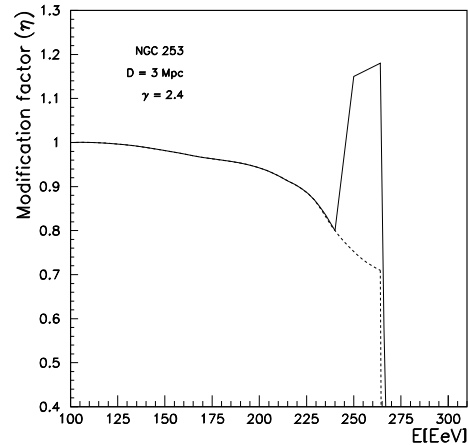}\\
\caption{Modification factor for initial iron nuclei from NGC 253,
assuming a differential power law 
injection spectrum with spectral index $\gamma = 2.4$. Solid (dotted) line 
stands for the case with an upper cutoff at the injection energy of 560 EeV
(340 EeV).}
\end{figure}

Events from the pile--up are about 50\% more probable to be detected
than those at energies immediately lower. The reason is that nuclei
with energies $\agt 560$ EeV
would be almost completely photodisintegrated during their
journey to Earth, in such a way that the surviving fragments end at
energies of the pile--up, changing the relative detection
probabilities. Thus, we  have the interesting result that the
existence of a sufficient high cutoff (at Lorentz factors, say,
$\Gamma \agt 8 \times 10^9$)  in the source acceleration
mechanism favors the detection of events around 250 EeV from the
starburst galaxy NGC 253.

\section{Concluding remark}

Might heavy nuclei be primaries at the end of the cosmic
ray spectrum? If the spectrum extends over the
up-to-now observed energies, the answer will be certainly ``no''. 
Heavy nuclei with
energies above 200 EeV could not propagate for more than 10 Mpc.
Besides, Lorentz factors above $2 \times 10^{10}$ require
large astrophysical regions for acceleration where the relic photon
density would be 
sufficient to provide a significant loss mechanism for the nuclei (see
Fig. 1).  

However, if the highest cosmic ray energy is of the order of 300 EeV,
heavy nuclei accelerated in the nearby starburst galaxies (NGC 253 and M82)
can be present at the end of the spectrum. We have shown that these
nuclei, originated in the central regions of the galaxies, can be
accelerated without suffering catastrophic interactions in a two-step
process that involves supernova remnant shock waves, and the large scale
terminal shock produced by the superwind that flows from the starbursts.   

Anchordoqui {\it et al.} \cite{prd2} have suggested that the
lack of data at energies immediately lower than the two 
``super--GZK events'' recorded to date, could be the
result of a different primary composition of disintegrated iron nuclei 
at the end of the spectrum. This speculation might find some support from 
our calculations since they suggest that the production of the energetic nuclei in
next-door galaxies (NGC 253 and M82) is at least feasible.  In any case, we 
have made some
concrete predictions that could be tested
with the forthcoming facilities of the Southern Auger Observatory
\cite{auger}, and the satellite experiment called OWL \cite{owl}.

\section*{Acknowledgements}

Several fruitful discussions on photonuclear interaction with M.T. Dova, 
L.N. Epele and E. Roulet are gratefully acknowledged. Especial thanks go
to D.F. Torres for a critical reading of the manuscript.
This work has been supported by FOMEC Program (L.A.A.), CONICET (through
research grant PIP 0430/98), and Antorchas Foundation (through funds
granted to G.E.R.).

\end{document}